\documentclass[epj]{svjour}
\usepackage{graphicx}
\usepackage{amssymb}
\begin{document}
\bibliographystyle{prsty}

\newcommand{\alps}{\ensuremath{\alpha_s}}
\newcommand{\qbar}{\bar{q}}
\newcommand{\beq}{\begin{equation}}
\newcommand{\eeq}{\end{equation}}
\newcommand{\beqa}{\begin{eqnarray}}
\newcommand{\eeqa}{\end{eqnarray}}
\newcommand{\gs}{g_{\pi NN}}
\newcommand{\gw}{f_\pi}
\newcommand{\mq}{m_Q}
\newcommand{\mn}{m_N}
\newcommand{\bb}{\langle}
\newcommand{\kb}{\rangle}
\newcommand{\st}{\ensuremath{\sqrt{\sigma}}}
\newcommand{\rvec}{\mathbf{r}}
\newcommand{\bvec}[1]{\ensuremath{\mathbf{#1}}}
\newcommand{\bra}[1]{\ensuremath{\bb#1|}}
\newcommand{\ket}[1]{\ensuremath{|#1\kb}}
\newcommand{\gft}{\ensuremath{\gamma_{FT}}}
\newcommand{\bfalp}{\mbox{\boldmath{$\alpha$}}}
\newcommand{\bfnab}{\mbox{\boldmath{$\nabla$}}}
\newcommand{\bfpi}{\mbox{\boldmath{$\pi$}}}
\newcommand{\bfsig}{\mbox{\boldmath{$\sigma$}}}
\newcommand{\bftau}{\mbox{\boldmath{$\tau$}}}
\newcommand{\spup}{\uparrow}
\newcommand{\spd}{\downarrow}
\newcommand{\hbarom}{\frac{\hbar^2}{m_Q}}
\newcommand{\half}{\frac{1}{2}}
\newcommand{\vnn}{\ensuremath{\hat{v}_{NN}}}
\newcommand{\argonne}{\ensuremath{v_{18}}}
\newcommand{\lqcd}{\ensuremath{\mathcal{L}_{QCD}}}
\newcommand{\lgf}{\ensuremath{\mathcal{L}_g}}
\newcommand{\lqm}{\ensuremath{\mathcal{L}_q}}
\newcommand{\lqg}{\ensuremath{\mathcal{L}_{qg}}}
\newcommand{\nn}{\ensuremath{NN}}
\newcommand{\hpnd}{\ensuremath{H_{\pi N\Delta}}}
\newcommand{\hpqq}{\ensuremath{H_{\pi qq}}}
\newcommand{\fpnn}{\ensuremath{f_{\pi NN}}}
\newcommand{\fpnd}{\ensuremath{f_{\pi N\Delta}}}
\newcommand{\fpqq}{\ensuremath{f_{\pi qq}}}
\newcommand{\ylm}{\ensuremath{Y_\ell^m}}
\newcommand{\ylmc}{\ensuremath{Y_\ell^{m*}}}
\newcommand{\qbh}{\hat{\bvec{q}}}
\newcommand{\xbh}{\hat{\bvec{X}}}
\newcommand{\dt}{\Delta\tau}
\newcommand{\qmag}{|\bvec{q}|}
\newcommand{\pmag}{|\bvec{p}|}
\newcommand{\oas}{\ensuremath{\mathcal{O}(\alpha_s)}}
\newcommand{\vtxb}{\ensuremath{\Lambda_\mu(p',p)}}
\newcommand{\vtxp}{\ensuremath{\Lambda^\mu(p',p)}}
\newcommand{\pwqp}{e^{i\bvec{q}\cdot\bvec{r}}}
\newcommand{\pwqm}{e^{-i\bvec{q}\cdot\bvec{r}}}
\newcommand{\gsa}[1]{\ensuremath{\bb#1\kb_0}}
\newcommand{\oer}[1]{\mathcal{O}\left(\frac{1}{\qmag^{#1}}\right)}
\newcommand{\nub}[1]{\overline{\nu^{#1}}}
\newcommand{\balph}{\mbox{\boldmath{$\alpha$}}}
\newcommand{\bgam}{\mbox{\boldmath{$\gamma$}}}
\newcommand{\epf}{E_\bvec{p}}
\newcommand{\epfp}{E_{\bvec{p}'}}
\newcommand{\eka}{E^\kappa_\alpha}
\newcommand{\ekaq}{(E^{\kappa}_\alpha)^2}
\newcommand{\ekap}{E^\kappa_{\alpha'}}
\newcommand{\ekpa}{E^{\kappa_+}_{\alpha}}
\newcommand{\ekma}{E^{\kappa_-}_{\alpha}}
\newcommand{\ekp}{E^{\kappa_+}}
\newcommand{\ekm}{E^{\kappa_-}}
\newcommand{\ekpap}{E^{\kappa_+}_{\alpha'}}
\newcommand{\ekmap}{E^{\kappa_-}_{\alpha'}}
\newcommand{\yjm}[1]{\mathcal{Y}_{jm}^{#1}}
\newcommand{\ysa}[3]{\mathcal{Y}_{#1,#2}^{#3}}

\title{Spacelike and timelike response of confined relativistic particles}
\author{Mark W. Paris}                     
\offprints{paris@lanl.gov}          
\institute{Theoretical Division, Los Alamos National Laboratory}
\date{Received: 13 September 2002 / Revised version: 13 September 2002}
%
\abstract{
Basic theoretical issues relating to the response of confined 
relativistic particles are considered including the scaling of the
response in spacelike and timelike regions of momentum transfer
and the role of final state interactions. A simple single particle 
potential model incorporating relativity and linear confinement 
is solved exactly and its response is calculated. The response
is studied in common approximation schemes and it is found that
final state interactions effects persist in the limit that the
three-momentum transferred to the target is large. The fact
that the particles are bound leads to a non-zero response in the
timelike region of four-momentum transfer equal to about 10\% of the
total strength. The strength in the timelike region must be taken 
into account to fulfill the particle number sum rule.
} 
\maketitle
\section{Introduction}
\label{intro}
Deep inelastic scattering (DIS) of leptons by hadrons is generally
discussed in the framework of the naive parton model and the
QCD-improved parton model using the operator product expansion.\cite{ESW}
This approach has been very successful in determining the
evolution of the structure functions as a function of the square
of the four-momentum transferred to the hadron.\cite{AP77}
In the leading order of the model the hadron is approximated by a 
collection of noninteracting quarks and gluons. 
The struck quark is assumed to be on the mass-shell
both before and after its interaction with the electron.
Basic theoretical considerations bring the validity of these 
assumptions into question.\cite{Bj00}

Based on the assumption that the struck constituent 
is on the mass-shell before and after interaction with the probe,
the response is predicted to be in 
the spacelike region for which the energy transfer $\nu$ is less
than the magnitude of momentum transfer, $\qmag$, as a consequence
of the inequality,
\beq
\label{eqn:nupwia}
\nu = \sqrt{|\bvec{k}+\bvec{q}|^2+m_q^2} - \sqrt{|\bvec{k}|^2+m_q^2}
\leq \qmag.
\eeq
Here $\bvec{k}$ and $m_q$ are the momentum and mass of the struck
quark, respectively.  The predicted response is discontinuous at 
the boundary $|\bvec{q}|=\nu$ between space and timelike regions. 
In fact interactions among the constituents in the initial state
take the constituents off the mass-shell and move response of the 
target into the timelike region of four-momentum transfer.

In the many-body theory (MBT) one expects,
at least naively, that final state interactions (FSI) should have an 
effect on inclusive scattering cross sections with electromagnetic 
probes from systems whose constituents are {\em confined}.
Scattering of high energy probes from composite systems, such as electron
scattering by nuclei \cite{Frois91} and nucleons \cite{ESW},
or neutron scattering by liquid helium \cite{SS89}, is often used to
study the structure of the bound system. The common assumption is that
in DIS at sufficiently high energy 
the probe is incoherently scattered by the constituents of the
system. In the plane wave impulse approximation (PWIA), which neglects
FSI effects, DIS is directly related to the momentum and energy 
distribution of the constituents in the target.

The role of FSI effects has been studied extensively in
electron scattering from nuclear targets \cite{BFFMPS91,BP93}
and neutron scattering from liquid helium \cite{SS89}. Recently
it has been suggested that they may also influence DIS of
leptons by hadrons \cite{Brodsky}.
In the present study we focus on scattering from targets with
confined constituents. The corresponding physical case concerns
DIS from nucleons where, in distinction from the nuclear and
liquid helium cases, the constituents are confined in both the
initial and final states.

\section{The response and scaling variables}
\label{sec:resp}
We consider the response to a hypothetical scalar probe which
couples to the density of a single scalar constituent. This allows
us to ignore complications due to spin and the Lorentz structure
of the response though it retains the qualitative features of
a more realistic model where one considers the coupling of a
spin-$\half$ fermion to the conserved electromagnetic current.
The response is
\beq
\label{eqn:rqn}
R(\bvec{q},\nu) = \sum_I |\bra{I} \sum_j e^{i \bvec{q}\cdot\rvec_j}
\ket{0}|^2 \delta(E_I - E_0 - \nu)
\eeq
where $\sum_j$ is over all the particles and the $\sum_I$
over all energy eigenstates. It is viewed
as the distribution of the strength of the state 
$\sum_j e^{i\bvec{q}\cdot\rvec_j}
\ket{0}$ over the energy eigenstates of the system having 
momentum $\bvec{q}$. It is
not necessarily zero in the timelike, $\nu > |\bvec{q}|$ region.

\subsection{Scaling variables}
\label{subsec:pmvar}
\begin{figure}[t]
\begin{center}
\includegraphics[ width=250pt, bb=0 0 640 640,
                  keepaspectratio, clip, angle=-90 ]{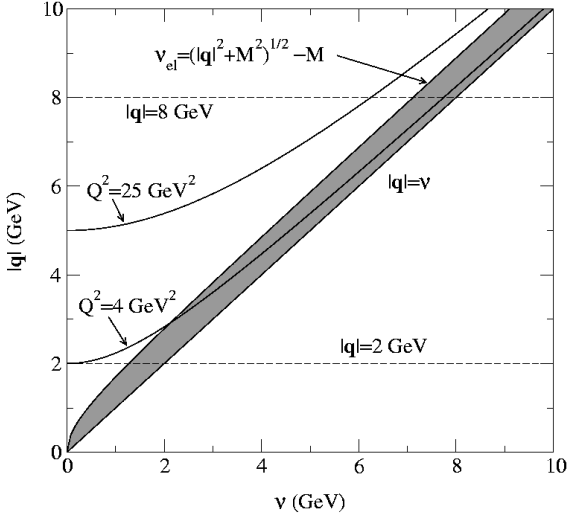}
\end{center}
\nopagebreak
\caption{\label{fig:qvn} The $\qmag$-$\nu$ plane. The spacelike region
is above the $\qmag=\nu$ line and timelike is below. Lines of constant
$Q^2 > 0$ are parabolas which lie entirely in the spacelike region
and approach $\qmag=\nu$ as $\nu\rightarrow\infty$. The observed
($Q^2>0$) response of the proton lies in the shaded area.}
\end{figure}

The conventional variables of the parton model,
$Q^2 = |\bvec{q}|^2 - \nu^2 $ and
the Bjorken $x=Q^2/2M\nu$, used to describe the DIS structure
functions of a hadron of mass $M$, are confined to the spacelike
region of the $\qmag$--$\nu$ plane for positive values of $Q^2$ 
accessible in lepton scattering experiments, as shown in
Fig.\ \ref{fig:qvn}. 
The observed ($Q^2>0$) DIS response is limited to a narrow region 
in the $\qmag-\nu$ plane illustrated in Fig.\ \ref{fig:qvn}. 
It is bounded by the elastic limit, $\nu_{el}=\sqrt{\qmag^2+M^2}-M$
on one side, and by the photon line on the other.  
In the limit of large $\qmag$ the width of the observed response 
at fixed $\qmag$ is $M$.
Lines of constant $Q^2$ intersect the elastic limit curve at
$x=1$ and approach the photon line at small $x$.

We wish to study the full range of response possible for a system
of bound constituents including the region of timelike momentum
transfer. Therefore we study the response,
$R(\bvec{q},\nu)$ as a function of $\nu$ and $\qmag$ in the rest
frame of the system \cite{BPS00}, as is common practice
in the MBT. Lines of constant $\qmag$ in Fig.\ \ref{fig:qvn} cross
the photon line ($\nu=\qmag$) and go into the timelike region.
The natural scaling variable in the MBT approach to 
DIS \cite{BPS00} is $\tilde{y}=\nu-|\bvec{q}|$. 
At large $|\bvec{q}|$ the response is expected to depend only on 
$\tilde{y}$, and not on $\bvec{q}$ and $\nu$ independently. This
variable is equivalent to the Nachtmann variable $\xi$ since
\cite{ON,Jaffe85}
\beq
\label{eqn:xi}
\xi = \frac{1}{M}(|\bvec{q}|-\nu) = -\frac{1}{M} \tilde{y}.
\eeq
In the limit of large $Q^2$ the 
$\xi = x$, thus $\tilde{y}$ scaling includes Bjorken scaling.  
However, both $\tilde{y}$ and $\xi$ span both spacelike 
and timelike regions at fixed $|\bvec{q}|$ unlike $x$ at fixed $Q^2$. 

\section{Model calculation}
\label{sec:model}
We have studied the exact response of a simple ``toy'' model
which contains the basic features of relativity and confinement
to obtain further insights on the possible response in the
timelike region and it's effects on the sum rules. In this model
we assume that the response of the hadron is due to a single
light valence quark confined within the hadron by its
interaction with an infinitely massive color charge. We model
this interaction by a linear flux-tube potential, and use the
single particle Hamiltonian,
\beq
\label{eqn:h1}
H = \sqrt{|{\bvec{p}}|^2+m_q^2} + \st \ r
\eeq
containing the relativistic kinetic energy operator. 
In the limit $m_q=0$ used here, the $H$ can be cast in the form:
\beq
\label{eqn:H}
H=\sigma^{1/4} \left( \sqrt{|\bvec{p^{\prime}}|^2} + r^{\prime} \right), 
\eeq
where $\bvec{p}'= \bvec{p}/\sigma^{1/4}$,
and $\bvec{r}'=\sigma^{1/4} \bvec{r}$ are dimensionless. 
The response $R(\qmag , \nu)$ of the model then depends only on the 
dimensionless variables $|\bvec{q}'|=\qmag /\sigma^{1/4}$ and 
$\nu'=\nu/\sigma^{1/4}$.  The main conclusions of this work are 
independent of the assumed value of $\sigma$; however, we show results 
in familiar units using the typical value $\sqrt{\sigma}=1$ GeV/fm. 

The model may be viewed as that of a meson with a heavy antiquark
or that of a baryon with a heavy diquark.  It is obviously 
too simple to address the observed response of hadrons.  For 
example, it omits the sea quarks and radiative gluon effects contained
in the DGLAP equations \cite{ESW,AP77} to describe scaling violations. 
Nevertheless its exact solutions are interesting and useful to study 
scaling, the approach to scaling, and the contribution of the timelike
region to sum rules.  A similar model has been considered by Isgur 
{\em et al.} \cite{Isgur01}. 

The Hamiltonian is diagonalized in the spherical momentum basis 
and the response is calculated to ensure
that the full strength of the integrated response, 
\beq 
\label{eqn:sumr}
\int_0^{\infty} R(\bvec{q},\nu) d\nu = 1, 
\eeq
is obtained in the chosen basis
for all values of the momentum transfer considered in this work
with $<$ 0.02 \% error.
In order to obtain a smooth response
we assume decay widths for all the excited states  
dependent on the excitation energy $\nu$.

\begin{figure}[t]
\includegraphics[ width=175pt, keepaspectratio, clip, angle=-90 ]{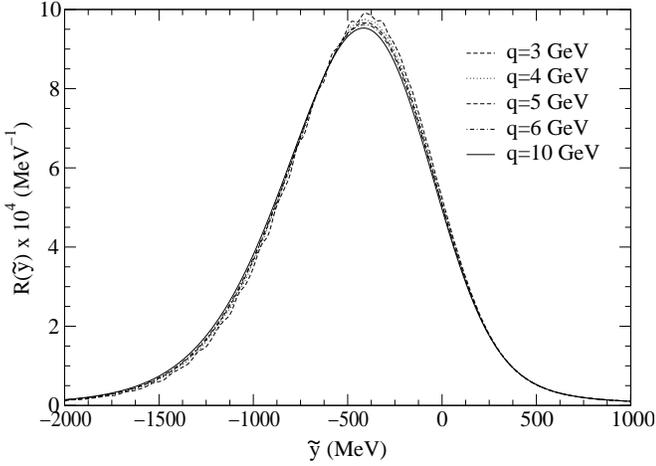}
\nopagebreak
\caption{\label{fig:rhq}
The response for values of $\qmag\geq 3$ GeV versus the scaling
variable, $\tilde{y}=\nu-\qmag$.}
\end{figure}

Figure \ref{fig:rhq} shows the response calculated for values of
$\qmag\geq 3$ GeV as a function of $\tilde{y}$.
The scaling behavior is clearly exhibited; at large $\qmag$ the 
$R(\qmag,\nu)$ becomes a function $f(\tilde{y})$ alone.  This 
scaling is equivalent to $\xi$ scaling via (Eq.~\ref{eqn:xi}).

\begin{figure}[t]
\includegraphics[ width=175pt, keepaspectratio, clip, angle=-90 ]{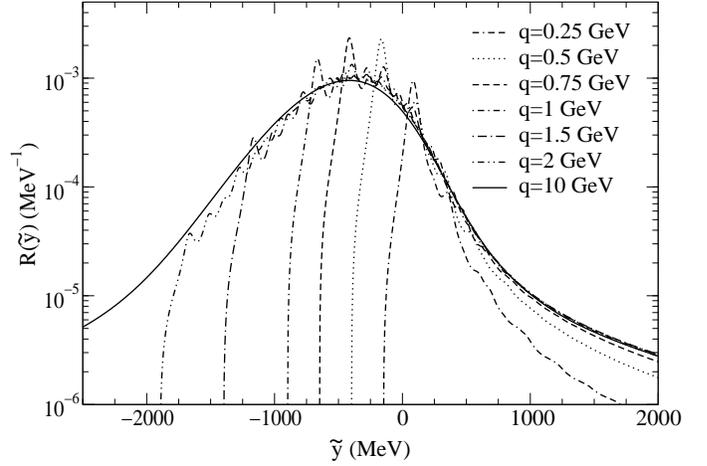}
\nopagebreak
\caption{\label{fig:ras}
The approach to scaling of the response for values of
$\qmag\leq 2$ GeV and $\qmag=10$ GeV
versus the scaling variable, $\tilde{y}=\nu-\qmag$.}
\end{figure}

In Fig.\ \ref{fig:ras} we show the response at 
various values of $\qmag\leq 2$ GeV compared with that for
$\qmag=10$ GeV, to study the approach to scaling.  At small $\qmag$ the 
scattering is dominated by resonances, and the first inelastic
peak is due to the lowest excited state with $n=1$ and $\ell=1$,
335 MeV above the ground state. 
In our toy model, the elastic scattering occurs at $\nu=0$ or
$\tilde{y}=-\qmag$, since our hadron is heavy. 
This elastic scattering
contribution is omitted from Fig.\ \ref{fig:ras}. 

For $\tilde{y} \sim 0$, {\em i.e.} for small $\xi$, 
the response approximately scales at relatively small values of
$\qmag$, comparable to $\sigma^{1/4}$. 
As $\qmag$ increases, the range over which scaling
occurs is extended to more negative 
values of $\tilde{y}$, {\em i.e.} to larger values of $\xi$. 
The contribution of each resonance shifts to lower $\tilde{y}$ 
and decreases in magnitude following the $R(\qmag \rightarrow \infty, 
\tilde{y})$. This behavior is
seen in the experimental data on the proton and deuteron 
\cite{Keppel} and interpreted as evidence for quark-hadron duality. 
Thus the toy model seems to describe some of the observed properties 
of the DIS response of nucleons. 
It exhibits $\tilde{y}$ or equivalently $\xi$ scaling 
at large $|\bvec{q}|$ as observed \cite{BPS00}, 
and an approach to $\xi$ scaling 
similar to that seen in recent experiments.

\subsection{Particle number sum rule}
In general the particle number sum rule in MBT is obtained by 
integrating the response at large $|\bvec{q}|$ over all $\nu > 0$:
\beqa
\int_0^{\infty}R(\bvec{q},\nu)d\nu &=& \sum_I \langle 0|\sum_i 
e^{-i\bvec{q}\cdot\bvec{r}_i} |I\rangle \langle I | \sum_j 
e^{i\bvec{q}\cdot\bvec{r}_j }|0\rangle   \nonumber \\
&=& \sum_{i,j} \langle 0|e^{i\bvec{q}\cdot (\bvec{r}_j-\bvec{r}_i)}|0\rangle.
\eeqa
When $\bvec{q}$ is large only the $i=j$ terms in the above
sum contribute, and therefore the integral gives the number
of particles in the system. In contrast the sums of the response
in the parton model are obtained by integrating the response
over $\xi > 0$ at fixed $Q^2$.  These sums will fulfill the
particle number sum rule only if the
response in the timelike region is zero. As mentioned earlier,
the response of a collection of noninteracting particles lies in
the spacelike region. Interaction effects, however, can shift a
part of the strength to the timelike region. Evidence for shifts
caused by interactions is discussed in Ref.\ \cite{BPS00}.

Returning to the ``toy'' model
the $R(\qmag,\nu)$, and therefore the $f(\tilde{y})$ 
extend into the timelike ($\tilde{y} > 0$) region. 
The sum-rule given by Eq.(\ref{eqn:sumr}), 
counts the number of particles in the target. It is 
necessary to integrate over the timelike region to fulfill
this sum rule. About 10\% of the sum is in that region 
independent of $\st$. The response
expressed as $R(Q^2,\xi)$ also scales at large $Q^2$ where
$|\bvec{q}|$ is necessarily large. It becomes a function of
$\xi$ alone. However, the integral:
\beq
\int_0^{\infty} R(Q^2 \rightarrow \infty,\xi)d\xi = \int_0^{|\bvec{q}|} 
R(\qmag \rightarrow \infty, \nu) d\nu \lesssim 0.9,  
\eeq
because the contribution of the timelike region is omitted. 
Here we have defined $\xi = \qmag - \nu $ without the conventional 
$1/M$ scale [Eq.(\ref{eqn:xi})].  

\section{Final state interaction effects}
\label{sec:fsi}
We study the effects of the FSI of the struck particle on the 
response. Analytic calculations of the width of the response are
presented for a general spherically symmetric potential and 
numerical results for a linear confining potential 
are  given. These indicate that the FSI increase the width of the
response beyond that predicted by PWIA. The analytic calculations 
also consider the nonrelativistic problem, in which $\bvec{q}$ is 
large compared to all the momenta in the target, but smaller than 
the constituent mass $m$. The main differences between the
nonrelativistic and the relativistic response are that the
former peaks at $\nu=\qmag^2/2m$ and has a width proportional
to $|\bvec{q}|$, while the latter peaks at $\nu \sim |\bvec{q}|$,
and has a constant width in the scaling limit.

\subsection{Moments of the response}
\label{subsec:mom}
In the case of a single confined particle, the state of the
system after the probe has struck the target is
\beq
\label{eqn:X}
\ket{X} = \pwqp\ket{0} ,
\eeq
where $\ket{0}$ denotes the ground state of the particle. The state
$\ket{X}$ is not an eigenstate of the Hamiltonian and therefore has
a distribution in energy. It  has a unit norm,
$\bb X|X \kb = \bra{0}\pwqm\pwqp\ket{0} = 1$.
The total strength of the response, given by the static
structure function
\beq
S(\qmag) = \int_0^{\infty} d\nu\, R(\qmag,\nu),
\label{eqn:S}
\eeq
is therefore unity.
In many-body systems $S(\qmag)$ is not necessarily equal to one.
Subsequent formulas pertain to the general case
and show factors of $S(\qmag)$ explicitly.

The mean
excitation energy of the state $\ket{X}$ is given by the first
moment of the response:
\beq
\label{eqn:barnu}
\overline{\nu}(\qmag) = \frac{1}{S(\qmag )} \bra{X} H-E_0 \ket{X}
=\frac{1}{S(\qmag)}\int_0^\infty d\nu\,\nu\, R(\qmag,\nu).
\eeq
The width of the distribution in energy is characterized
by the second moment of the energy about the mean:
\beq
\label{eqn:delta}
\Delta^2(\qmag) = \frac{1}{S(\qmag)}\bra{X}
\left( H - \frac{\bra{X}H\ket{X}}{S(\qmag)} \right)^{\!2} \ket{X}.
\eeq

Substitution of the Eq.(\ref{eqn:X}) into the formulas for the
first three moments of the response give the following results:
\beqa
\label{eqn:barnux}
\overline{\nu}(\qmag) &=& \qmag + \gsa{V} - E_0 \nonumber \\
&+& \frac{1}{3\qmag}\gsa{k^2} + \oer{3}; \\
\label{eqn:delgs}
\Delta^2(\qmag) &=& \frac{1}{3}\gsa{k^2} + \gsa{V^2} - \gsa{V}^2 \nonumber \\
&+& \frac{2}{3\qmag}\left(\gsa{k^2V}-\gsa{k^2}\gsa{V}\right)
+ \oer{2}.
\eeqa
Here $\oer{n}$ denotes the neglected terms of that and higher order
and the angle brackets with subscript `0' indicate averaging with
respect to the ground state.
Thus $\overline{\nu}(\qmag)= \qmag - \gsa{T}$ in the limit $\qmag 
\rightarrow \infty$, where $T=\sqrt{|\bvec{p}|^2}$ is the kinetic energy.
The requirement that $\overline{\nu}(\qmag)-\qmag$ becomes constant
is naturally satisfied in this limit.
These expression demonstrates that the average energy and 
width of the exact response
is independent of $\qmag$ in the limit $\qmag\rightarrow\infty$
as necessary for $\tilde{y}$ scaling. It also shows that the width
has a kinematic contribution dependent upon the target
momentum distribution, and an additional interaction contribution.

As mentioned, the PWIA assumes that a constituent of momentum $\bvec{k}$,
after being struck by the probe, may be described
by a plane wave with momentum $\bvec{k+q}$ in an assumed
average potential chosen to give the exact $\overline{\nu}$
of Eq.(\ref{eqn:barnux}). From the PWIA response we calculate
\beq
\label{eqn:delpw}
\Delta^2_{PWIA} = \frac{1}{3}\gsa{k^2} + \oer{2}
\eeq
contains only the first term of the exact result [Eq.(\ref{eqn:delgs})]
due to the target momentum distribution.  The second term,  
$\gsa{V^2}-\gsa{V}^2$ of Eq.(\ref{eqn:delgs}) represents
the FSI contribution neglected in the PWIA.  It does not vanish in the
$|\bvec{q}| \rightarrow \infty$ limit for relativistic kinematics.

In the non-relativistic case,
$H_{NR} = \frac{|\bvec{p}|^2}{2m} + V(r)$,
the exact $\overline{\nu}$
is given by:
\beq
\label{eqn:barnupwnr}
\overline{\nu}_{NR} = \frac{1}{2m}\qmag^2 \ .
\eeq
For the width of the NR-PWIA response we obtain:
\beqa
\label{eqn:delpwnr}
\Delta^2_{NR-PWIA}(\qmag) &=& \qmag^2 \frac{\gsa{k^2}}{3 m^2} \nonumber \\
&+&\frac{1}{4m^2} \left( \gsa{k^4} - \gsa{k^2}^2 \right).
\eeqa
Note that in Eqs.(\ref{eqn:barnupwnr}) and (\ref{eqn:delpwnr})
we have {\em not} taken the $\qmag\rightarrow\infty$ limit.

The width of the exact NR response is:
\beqa
\label{eqn:delnr}
\Delta^2_{NR}(\qmag) &=& \Delta^2_{NR-PWIA}
+ \frac{1}{m}\left(\gsa{Vk^2} - \gsa{V}\gsa{k^2}\right) \nonumber \\
&+& \gsa{V^2} - \gsa{V}^2 \ .
\eeqa
It differs from $\Delta_{NR-PWIA}$ in terms of
order $1/|\bvec{q}|$ which can be neglected in the
scaling limit.  Thus, in contrast to the relativistic case,
the FSI do not increase
the width of the NR-PWIA response at large $|\bvec{q}|$.

Finally we consider the on-shell approximation (OSA) in
which the energy of
the struck constituent is that of a free relativistic
particle before and after the interaction with probe,
as assumed in the quark-parton model. The response in OSA
depends only on the momentum distribution of target constituents
and obeys $\tilde{y}$ scaling. The average excitation in OSA is
\beq
\label{eqn:barnuosa}
\overline{\nu}_{OSA} = \qmag -\gsa{T} + \frac{1}{3\qmag}\gsa{k^2}
+\oer{3},
\eeq
and the width is given by:
\beq
\label{eqn:delosa}
\Delta^2_{OSA} = \frac{1}{3}\gsa{k^2} + \gsa{k^2}-\gsa{T}^2
+ \oer{}.
\eeq
The exact value of $\overline{\nu}$ [Eq.(\ref{eqn:barnux})]
is reproduced by the OSA for any potential.  However, the
$\Delta^2_{OSA}$ has $\gsa{k^2}-\gsa{T}^2$ in place of the
$\gsa{V^2}-\gsa{V}^2$ in the leading term of the exact $\Delta^2$
[Eq.(\ref{eqn:delgs})].
For a massless particle in a linear confining potential,
{\em i.e.}\ for the Hamiltonian of Eq.(\ref{eqn:H}),
$\gsa{T}=\gsa{V}$, and $\gsa{k^2}=\gsa{V^2}$. Therefore for
this particular Hamiltonian the OSA reproduces the
exact value of $\Delta$; but the shape is wrong.

\begin{figure}
\begin{center}
\includegraphics[ width=225pt, keepaspectratio, angle=0, clip ]{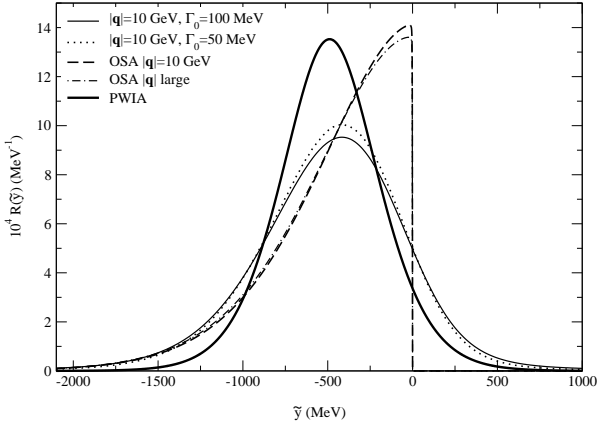}
\end{center}
\caption{\label{fig:rcf} The response versus $\tilde{y}$ calculated
exactly for $\Gamma_0=100$ MeV (thin solid curve) and $\Gamma_0=50$ MeV
(dotted curve). The response in OSA are shown for $\qmag=10$ GeV (dashed)
and $\qmag\rightarrow\infty$ (dot-dashed). The PWIA response for
$\qmag=10$ GeV and $\qmag\rightarrow\infty$ lie on essentially the
same (thick solid) curve.}
\end{figure}

\subsection{Numerical results}
\label{subsec:num}
We first compare the response functions for $\qmag = 10 $ GeV
before comparing their moments. In Ref. \cite{Paris01} it has been
shown that the scaling limit is obtained for such values of
$\qmag$.  The exact response,
Eq.(\ref{eqn:rqn}), is a sequence of $\delta$ functions at
$\nu=E_I-E_0$. In order to obtain a smooth response
we assume decay widths $\Gamma_0$ for all the excited states. 
Note that the energies of the states $|I\rangle$ that contribute to
the response at $\qmag = 10 $ GeV are large, 
therefore their decay widths are not
affected by the energy dependent terms assumed in Ref. \cite{Paris01}.
The response including decay widths is given by:
\beq
\label{eqn:smear}
R(\bvec{q},\nu) = \sum_I|\langle I| e^{i \bvec{q} \cdot \bvec{r}}
|0 \rangle|^2 \left(\frac{\Gamma_0}{2\pi}\right) \frac{1}
{(E_I-E_0-\nu)^2 + \Gamma_0^2/4} \ .
\eeq
The responses obtained with $\Gamma_0=100$ and 50 MeV are shown in
Fig.\ \ref{fig:rcf}, along with the PWIA
and OSA responses for $\qmag = 10$ GeV and for $\qmag \rightarrow \infty$. 
The difference between the exact responses for $\Gamma_0=100$ and 50
MeV are much smaller than those between the exact and the approximate.

We note that the shape of the
PWIA response is qualitatively similar to that of the exact,
however, its width is too small.
This is a direct consequence of the neglect of interaction
terms in $\Delta$ [Eq.(\ref{eqn:delgs})] 
as discussed in the last section.  The width $\Delta$ of the response
is 409 MeV, while the $\Delta_{PWIA} = 326$ MeV. 

The OSA results in the discontinuous curves shown in
Fig.\ \ref{fig:rcf}. They are  discontinuous at the
lightline ($\qmag=\nu$) because the response of free
particles is limited to the spacelike region $\nu<\qmag$.
The discontinuity at $\tilde{y}=0$ is in clear conflict with
the exact response which is continuous across the lightline
and is non-zero in the timelike ($\tilde{y}>0$) region.
Therefore the OSA appears to be unsatisfactory even though for
the special case of a linear potential it
has the exact values of $S(\qmag)$, $\overline{\nu}(\qmag)$ and
$\Delta(\qmag)$.

\end{document}